\documentclass[12pt]{iopart}

\usepackage{graphics}
\begin{document}
\title [On the Capability Of Super-Kamiokande Detector]
{
On the Capability Of SuperKamiokande Detector To Define the Primary Parameters Of Muon And Electron Events
}

\author{
V.I.~Galkin$^1$, A.M.~Anokhina$^1$,
E.~Konishi$^2$ and A.~Misaki$^3$
}
\address{
$^1$Department of Physics, Moscow State University, Moscow, 119992, Russia\\
$^2$Graduate School of Science and Technology, Hirosaki University, Hirosaki, 036-8561, Japan\\    
$^3$Research Institute for Science and Engineering, Waseda University, Tokyo, 169-0092, Japan 
}
\ead{konish@si.hirosaki-u.ac.jp}

\begin{abstract}
  We develop a new discrimination
 procedure for separating electron neutrinos from muon neutrinos,
 based on detailed simulations carried out with GEANT~3.21 and
 with newly derived mean angular distribution functions for the
 charged particles concerned (muons and electrons/positrons),
 as well as the corresponding functions for the relative
 fluctuations.  These angular distribution functions are
 constructed introducing a ``moving point'' approximation.
 Using our type definition procedure we are able to discriminate muons from
electrons in Fully Contained Events with a probability of error of
less than several \%.  At the same time, our geometrical reconstruction
procedure, considering only the ring-like structure of the Cherenkov
image, gives an unsatisfactory resolution for 1~GeV $e$ and $\mu$,
with a mean vertex position error, $\delta r$, of 5--10\,m and a mean
directional error, $\delta \theta$, of about 6$^\circ$--20$^\circ$.
In contrast, a geometrical reconstruction procedure utilizing the full
image and using a detailed approximation of the event angular
distribution works much better: for a 1~GeV $e$, $\delta r \sim$2\,m
and $\delta \theta \sim$3$^\circ$; for a 1~GeV $\mu$, $\delta r
\sim$3\,m and $\delta \theta \sim$5$^\circ$.  At 5~GeV, the
corresponding values are $\sim$1.4\,m and $\sim$2$^\circ$ for $e$ and
$\sim$2.9\,m and $\sim$4.3$^\circ$ for $\mu$.  The numerical values
depend on a single PMT contribution threshold. The values quoted above
are the minima with respect to this threshold.
Even the methodologically correct approach we have adopted, based on
detailed simulations using closer approximations than those adopted in
the SK analysis, cannot reproduce the accuracies for particle
discrimination, momentum resolution, interaction vertex location, and
angular resolution obtained by the SK simulations, suggesting the
assumptions in these may be inadequate.
\end{abstract}

\maketitle

\section{Introduction}
\label{intro}
The possible existence of neutrino oscillations is one of the most 
important issues in particle astrophysics as well as elementary 
particle physics at the present time. 
Among the positive and negative results reported for neutrino 
oscillation, experimental results for atmospheric neutrino by 
Super Kamiokande (hereafter, we abbreviate simply SK) has special position 
in the experiments concerned, because it is said that they have 
given the decisive and clear evidence 
for the existence of neutrino oscillation.
The reasons as follows: \newline
\indent (1) They carried out the calibration experiments 
for the discrimination between muon and electron by electron 
accelerator beam whose energies are well known and established 
the clear discrimination between muon and electron 
for SK energy region concerned \cite{Kasuga1}. \par
  
(2) Based on the well established discrimination procedure between 
muon and electron, they have analyzed {\it Fully Contained Events} 
and {\it Partially Contained Events}, whose energies covered from 
several hundreds MeV to several GeV.  As the results of them, 
they have found significantly different zenith angle distribution 
for muon and electron, namely, muon deficit, and 
attributed such discrepancy to  the neutrino oscillation between 
$\nu_{\mu}$ and $\nu_{\tau}$.
As the most new one, they give   ${\rm sin}^2{2\theta} > 0.92$ and 
$1.5\times 10^{-3}{\rm eV}^2 < \Delta m^2 < 3.4\times 10^{-3}{\rm eV}^2$ 
at 90\% confidence level \cite{Fukuda1}.
 \par
 (3) Also, they have analyzed {\it Upward Through Going Particle Events} 
 and {\it Stopping Particle Events}. Most physical events under such 
 category could be regarded as exclusively  the muon (neutrino)
induced events, not electron (neutrino) induced events, because the 
effective volume for muon is much larger than that for electron due 
to longer range of muon irrespective of the discrimination procedure 
between muon and electron which is indispensable for the analysis for 
{\it Fully Contained Events} and {\it Partially Contained Events}.  
Also, in this case, they have given the same parameters for neutrino 
oscillation which are obtained in the analysis of 
{\it Fully Contained Events} 
and {\it Partially Contained Events} \cite{Fukuda1}. 

Through three different kinds of the experiment performed by SK, 
all of which are constructed upon the well established procedure, 
it is said that SK has given clear and definite evidence for existence
 for the neutrino oscillation.
 \par
  
 The analysis of {\it Fully Contained Events} and {\it Partially Contained 
 Events} is closely and inevitably related to the discrimination 
 procedure between electron and muon. 
The frequency of muon  events with some energy 
occurred inside the detector is nearly the 
 same as that of electron events unless ossillation exists and,
 therefore, the precise discrimination procedure between
 electron and muon is absolutely necessary.

   Considering the great impact of SK experiment over other experiments 
concerned and theoretical physics, we feel we should examine the 
validities of the experimental results performed by SK, because 
nobody has examined them in the most comprehensive way, solely 
due to character of huge experiment, although the partial aspect 
of SK had been examined in fragmental way \cite{Olga}.

   However, Mitsui et al have examined the validity of the 
discrimination procedure by SK and have pointed out the necessity 
of fluctuation effect into the discrimination procedure 
between muon and electron  by SK \cite{Mitsui}.

   We have examined validities of all the SK experiment, adopting 
quite different approach from the SK procedure.

\section{Algorithm For Processing Cherenkov Light Images In SuperKamiokande Experiments}
\label{sec:2}
                                                        
    As can be inferred from \cite{Sakai,Kasuga3}, the image processing technique
at SuperKamiokande is based on events simulated with the aid of the GEANT3.21 code \cite{GEANT}.

 But, in reality, a small part of the simulation results is used by SK to construct models of $e,\mu$-event
images, namely, the average angular distribution (more precisely, that which is averaged both over
shower particles and over the ensemble of showers) of light emitted from the electromagnetic
shower initiated by an electron. In the SuperKamiokande studies, the spatial distribution of the
light source (that is, the shower) is not taken into account either in the transverse direction or
along the shower axis --- in other words, the shower is taken in the pointlike form. This may lead
to significant distortions of the pattern in procedures for event-type recognition and event-geometry
reconstruction, since the mean longitudinal length of a shower initiated by particles of energy
1 GeV is about 4 m, while the dimensions of the sensitive volume of the water tank do not exceed 40 m,
the events being uniformly distributed over the whole tank volume. Further, a muon track is represented
by a straight-line segment, the distribution of light emitted from it taking the form of a delta
function (that is, the photons fly along the Cherenkov cone generatrices). This means that the effect
of multiple scattering is neglected in \cite{Sakai,Kasuga3}.

      The disregard of the information about fluctuations of the light spatial and angular distributions
that is contained in simulated events is yet another significant simplification that is not well
justified in our opinion. The scale of relative fluctuations of the light angular distribution for
events initiated by particles of energy about 1 GeV is about hundreds of percent, and it is this
circumstance that must restrict substantially the potential for the reconstruction of the event type
and geometry.

    Mitsui et al. \cite{Mitsui} also indicated that the event models chosen by the SuperKamiokande Collaboration
are inadequate. They reproduced the SuperKamiokande procedure for events obtained from a Monte Carlo
simulation with allowance for all possible fluctuations (including fluctuations of photoelectrons) and
found errors in event-type identification that are much greater than those reported by the SuperKamiokande
Collaboration (about 20\% for events initiated by particles of energy below 1 GeV versus several percent).

\section{Statement Of the Problem Of Estimating the Upper Limits For the Parameter Resolutions}
\label{sec:3}

    Here, we did not aim at developing new algorithms for processing SuperKamiokande data, since this
would require considerable resources and detailed knowledge of the experimental setup; instead, we just   
tried to set absolute limits on the potential of SuperKamiokande telescope that    
are associated with detector geometry and physical processes of light generation and propagation.          

    The problem of determining primary parameters of events can be simplified by breaking it down into three
separate problems:

    (i) that of determining the primary-particle momentum (energy) under the assumption that the particle
type, the injection point (particle-production vertex), and the direction of particle motion are known;
                                                        
    (ii) that of identifying the primary-particle type under the assumption that the particle momentum
(energy), the injection point, and the direction of particle motion are known;
                                                        
    (iii) that of determining injection point for a primary particle and the direction of its motion under
the assumption that its type and momentum (energy) are known.
                                                        
    For all of the parameters, this approach is generally expected to give resolutions that are higher
than those in the case of solving the total problem of determining all parameters simultaneously. Thus,
our results must set limits on the resolutions of the SuperKamiokande telescope.

\begin{figure*}[t]
\rotatebox{0}{ \resizebox{0.93\textwidth}{!}{\includegraphics{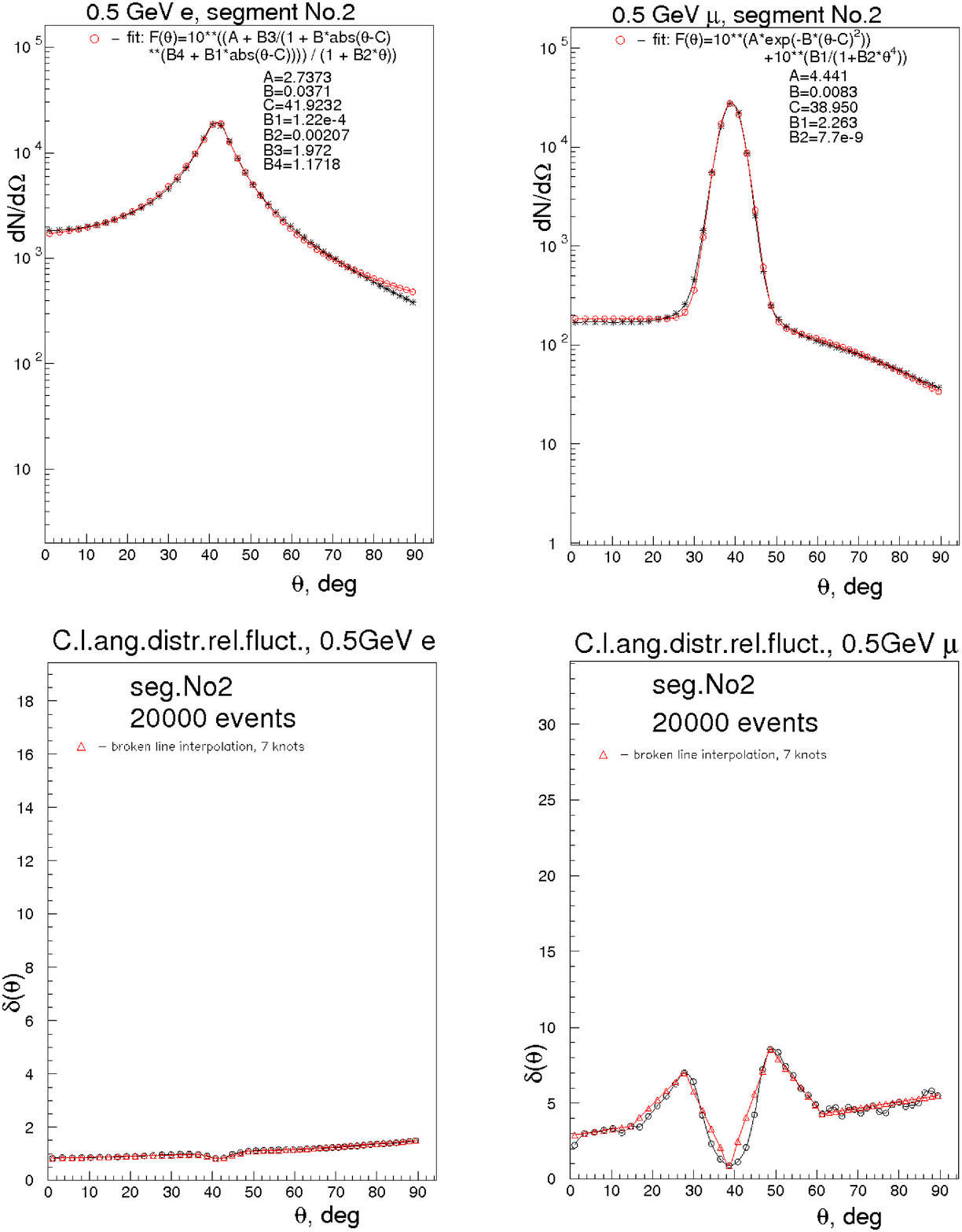}}}
\caption{Mean angular distribution ($dN/d\Omega$) of Cherenkov light from 0.5 GeV electron and muon and the
corresponding relative fluctuations. The data obtained for segment 40-80 $g/cm^2$. Samples used: 20000 events.}
\label{ang}
\end{figure*}

    In the case where a muon track or an electron shower lies completely within the sensitive volume
of the detector (so-called Fully Contained Events), a precise description of event geometry, especially
in the SuperKamiokande detector, where the light absorption length is about 100 m, is not required for
the first problem because the total amount of recorded Cherenkov light depends strongly on the primary
particle momentum.

In this paper we limit our consideration to the analysis of Fully Contained Events.
                                                        
    As for the second and third problems, it is necessary to describe in detail the features of Cherenkov
light emitted by a muon and an electron --- at least in three dimensions --- since, in order to determine
the event type and event geometry, one needs the image pattern to compare with it the actual experimental
image. An oversimplified description of images can impair considerably the resolution in the event type
and event geometry.

     The process of optical-photon transformation into photoelectrons was not considered, because its
analysis would require detailed information about photomultiplier tubes. This simplification does not
change the initial purpose of estimating limits on resolutions, since the elimination of one source of
fluctuations may only improve the respective estimates.

     We decided to neglect the lateral distribution of particles in electron and muon events (its scale was
on the order of a few tenths of a meter); that is, we assumed that Cherenkov light is emitted exclusively
from the event axis. In view of this, the present approach is inapplicable to events associated with
particles moving overly close to the tank walls and nearly parallel to them, but it is quite suitable for
estimating resolutions.

\section{Monte-Carlo Simulation}
\label{sec:4}

     In simulating images, we assumed that events occurred within a cylindrical water volume 16.9 m in
radius and 36.2 m in height. Further, 11 408 photomultiplier tubes, the diameter of their photocathodes
being 50 cm, were distributed uniformly over the walls of the cylinder and over its bottom and top bases.
This corresponds roughly to the SuperKamiokande detector. The scattering of light and its reflection
from the walls were disregarded, but its absorption was taken into account.

     A modified GEANT3.21 code was used to obtain simulated Cherenkov light images for events in the above
discribed analog of the SuperKamiokande telescope and to develop an adequate model of muon and electron
events. In our modification of the code, we abandoned the standard algorithms for tracking optical photons.
Instead, the product photon was tracked along a straight line until it hit the wall and was considered to
be recorded with the weight equal to the transmission coefficient for the traversed water layer if the
trajectory of this photon intersected a circle imitating a photomultiplier tube.

     The water refraction index was taken to be 1.34 for the whole wavelength range 300-600 nm considered
here. Electrons were tracked up to the kinetic energy of 0.25 MeV, while muons were tracked up to their decay.
As a result of event simulation one gets detailed Cherenkov images of $e$-showers/$\mu$-tracks in SK telescope.

For the construction of event parameter definition procedures
one needs the models of event images that are close
to real ones. Event parameters are defined through a comparison
of these model images with different parameter
values with experimental images.

Instead of the simple mean models used by SK (point-like one
for $e$-events and straight line for $\mu$-events) we
introduce 'moving-point' approximation models for both $e$
and $\mu$-events. In this approximation
$e$-shower/$\mu$-track is assumed to emit Cherenkov
photons from a straight line following the primary direction
but the angular distribution of light changes along this line
i.e. both mean angular distribution and its
fluctuation evolve with water layer depth.

     The procedure for simulating electron and muon light angular distribution involved the following steps:

     (i) The mean longitudinal length of a muon track (electron shower) was fixed to be the length from
which 99.5\% of the mean total number of Cherenkov photons are emitted.


    (ii) This length was broken down into equal segments, their length and number depending on the particle
type and energy and on the required accuracy in the image pattern.
                                                                  
    (iii) The mean angular distribution of Cherenkov light, $F_i^{e,µ} (\theta)$, and its relative fluctuation
$\delta_i^{e,µ} (\theta)$ were calculated for each segment $i$.
                                                                  
    Thus, the simulation provides an approximation of the mean angular distributions of light
    
\begin{equation*}
F_{e,\mu} (\theta_i,E_0,k) \, = \, \frac{\left< N_{e,\mu}(\theta_i,E_0,k)\right >}
{\Delta\Omega_i},
\label{eqn:12}
\end{equation*}
\vspace{0.5cm}
and their relative fluctuations\\

\begin{equation*}
\delta_{e,\mu} (\theta_i,E_0,k) \, = \, \frac{\sqrt{\left<N_{e,\mu}^2(\theta_i,E_0,k)
\right> \,- \, \left<N_{e,\mu}(\theta_i,E_0,k)\right>^2}}
{\left<N_{e,\mu}(\theta_i,E_0,k)\right>} \;.
\label{eqn:13}
\end{equation*}
\vspace{0.5cm}
versus the light emission angle $\theta$ and the water-layer thickness $t$. Here $\left<...\right>$
denotes the  average over a large event sample, $N_{e,\mu}(\theta_i,E_0,k)$ is a number of Cherenkov
photons emitted from segment $k$, $\theta_i$ is the center of mass of the $i$-th histogram bin, and 
$\Delta\Omega_i$ is the solid angle of the $i$-th bin.
 While calculating $F_{e,\mu} (\theta_i,E_0,k)$ and $\delta_{e,\mu} (\theta_i,E_0,k)$ we considered
samples of 10 000 to 20 000 events and did not track Cherenkov               
photons, but we included their contributions in the histograms
in $\theta$ of bin width 1.875${}^o$ immediately after
light generation, irrespective of the azimuthal emission angle.
                                                                  
    The number of the segments was varied from 7 to 24 in the
calculations; segments of length 40 cm and 100 cm were
used for events generated by particles of energy below 1 GeV and equal to 5 GeV, respectively.              

    In order to approximate the mean angular distribution of
light within each individual segment of a muon track
(electron shower), we took the model functions\\
\begin{equation*}                                                           
F_{\mu}(\theta;A,B,C,B1,B2) = 10^{\left \{ A \, \mathrm{exp}
 \left [ -B \, (\theta - C)^2 \right ]
\right \}} \, + \,  10^{\left [ B1/(1+B2 \, \theta^4)
\right ]} 
\label{eqn:14}
\end{equation*}
\begin{equation*}
F_e(\theta;A,B,C,B1,B2,B3,B4) = = 10^{\left \{
\frac{ A \, + \, B3/ \left ( 1 + B \cdot |\theta
- C|^{\left ( B4 + B1 \cdot |\theta - C| \right )}
\right )}{1 \, + \, B2 \cdot \theta} \right \}} 
\label{eqn:15}
\end{equation*}
\vspace{0.5cm}
The approximations of the mean angular distributions were obtained as the best least squares fits of the
model functions to the histograms.                                

    The shapes of the relative fluctuations are very complicated.
However, we used linear interpolations with 7 to
8 nodes to describe the fluctuations, because a high accuracy was not necessary in that case.

      Figure 1 shows examples of approximations of the mean angular distributions of light and their relative
fluctuations in events initiated by electrons (a, b) and muons (c, d) of energy 500 MeV.

      It should be noted that, for a given type and a given
energy of the primary particle, the above approximations
of the angular distributions of Cherenkov light are quite
universal in the sense that they can be used to calculate
the patterns of mean images and their variations for any
possible geometry of events in any water detector.

\section{Procedures For Reconstructing Primary Parameters Of Events}
\label{sec:5}

\subsection{Reconstruction of Primary Energy (Momentum)}

      As was mentioned above, the energy for fully contained events
of a specific type can be estimated on the basis
of the total number of photons recorded by photomultiplier tubes.
 Therefore, the energy (momentum) resolution can be
estimated from the width of the distribution of the total number
of recorded Cherenkov photons. For muons and electrons
of momentum 300 MeV/c emitted approximately from the tank center,
Fig. 2 shows the distributions of the total number
of recorded photons. The relative fluctuations obtained in
this study for events initiated by electrons and muons having
two different momenta are presented in Table 1, along with the
estimates of the momentum resolution from \cite{Kasuga3}.

\begin{figure}
\begin{center}
\rotatebox{0}{ \resizebox{0.45\textwidth}{!}{\includegraphics{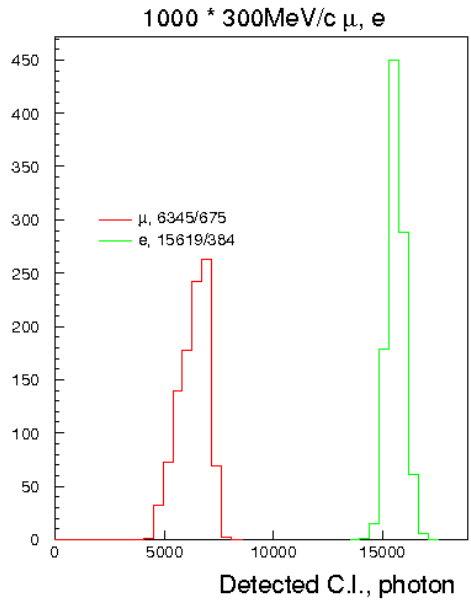}}}
\caption{Distribution in the total number of detected Cherenkov photons from muon and electron events
initiated by 300 MeV/c particles. The injection point is in the center of the tank. Sample volume:
1000 events.}
\label{totdet}
\end{center}
\end{figure}

\begin{table}
\caption{Energy (momentum) determination errors} \label{t1}
\begin{center}
\begin{tabular}{|c|c|c|c|c|}
\hline
momentum, MeV/c &  \multicolumn{2}{|c|}{300} & \multicolumn{2}{|c|}{1000} \\
\hline
event type & $\mu$ & $e$ & $\mu$ & $e$   \\
\hline
present work & 10,6 & 2,5 & 1,9 & 1,3 \\
\hline
\cite{Kasuga3} & 3,0 & 5,3 & 2,4 & 3,2 \\
\hline
\end{tabular}
\end{center}
\end{table}

   That fluctuations in muon events are much larger than those
in events initiated by electrons is quite understandable:
muons of energy about 1 GeV lose energy only by ionization;
in a muon event, one particle carries the bulk of the energy,
and the fluctuations of the total number of recorded photons
reflect a wide diversity of possible muon-propagation histories.
In an electromagnetic shower, the energy is distributed among
many particles, with the result that fluctuations of its
features are less pronounced.

      As the muon momentum decreases from 1 GeV/c to 300 MeV/c,
fluctuations of the total number of emitted photons
increase considerably, which is due to an increase in the
relative contribution to Cherenkov radiation from that portion
of muon events which experience the greatest fluctuations:
the total number of photons depends on the location of the
decay vertex along the track and on the energy of the decay
electron, which also emits light.

           The collection of light by detectors introduces additional
uncertainties since only about one-third of
the tank-wall area is covered with photomultiplier tubes and
since the distribution of photons over this area is governed
by the fluctuating spatial and angular distribution of light in the source.

     It can be stated that the resolutions in electron events are in
reasonable agreement (if we disregard fluctuations of
photon transformation into photoelectrons) --- that is,
the uncertainties obtained in our study are smaller than those in \cite{Kasuga3}.
 At the same time, the resolution values for muons disagree,
at least for momenta below 1 GeV/c.                        
                                                        
     There is a characteristic relationship between the total
amount of light generated in muon and electron    
events at identical momenta of primary particles:
the number of photons generated by a muon is smaller by     
$2.8 \times 10^4$ than the number of photons generated by an
electromagnetic shower, this being due to the difference in the
Cherenkov thresholds. It is necessary to take this fact into account in
addressing the problem of event-type identification.
It is logical to formulate this problem for events involving identical
 numbers of recorded Cherenkov photons. For the
sub-GeV and GeV energy ranges considered here, this means that
one should consider muon events of energy about   
200 MeV higher than the energy of electron events.

\subsection{Event-Type Identification}

    In the problems of event-type identification and the reconstruction of event geometry, each event is
treated as some random vector ${\bf Q} = {Q_j}$ whose components are the
 contributions (numbers of Cherenkov photons) to all
photomultiplier tubes of the setup. Here, j is the photomultiplier
 index (j = 1, 2, . . ., N ) and N is the number of
photomultiplier tubes. In the procedure of identifying the event type, one considers two classes of events:
$\omega_1$ = e (electron) and $\omega_2$ = $\mu$ (muon). A Monte Carlo
 simulation of event optical images makes it possible
to study the properties of images belonging to both classes --- that is, to obtain the image distributions
$F (Q_1 , Q_2 , . . . , Q_N ; \omega_i, E_0, {\bf r}_0, {\bf \theta}_0)$,
 which are the joint distributions of the light
contributions $Q_j$ to photomultiplier tubes for the case where the particle
 type $\omega_i$, the particle energy $E_0$,
the particle injection point ${\bf r}_0$, and the quantity ${\bf \theta}_0$
 specifying the direction of particle motion are preset.

    However, it is hardly possible to deal with such functions in practice,
     because one has to simulate a great event sample
in order to obtain a distribution function that involves this extent of differentiation for each set
${\omega_i , E_0 , {\bf r}_0 , {\bf \theta}_0 }$.

    In order to construct a more realistic solution to this problem,
it would be more appropriate to choose
an adequate model of the distribution of the number of Cherenkov photons
in an individual detector and to specify such
a distribution in each individual photomultiplier tube in terms of
only the first few of its moments. In the case being
considered, these distributions are close to a normal
distribution at rather large mean numbers of photons (Fig. 3).
\footnote{In real cases, the distributions of features used for
classification bear most often much less resemblance to normal
distributions than the distributions in Fig. 3, but this
does not prevent their approximation by a normal distribution
for multidimensional classification. As a matter of fact,
it is necessary that the distribution density for the features being
considered have one maximum, not overly large asymmetry,
and two first momenta. With increasing dimensionality of the
feature vector, higher order moments become progressively less significant.}

\begin{figure}
\begin{center}
\rotatebox{0}{ \resizebox{0.45\textwidth}{!}{\includegraphics{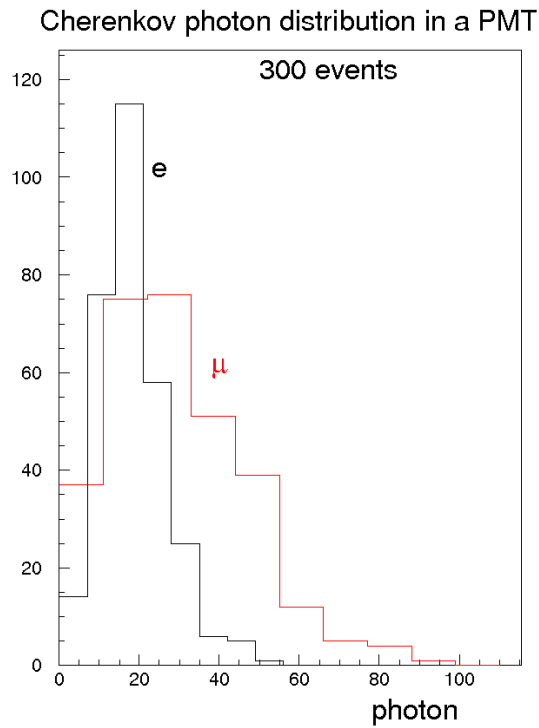}}}
\caption{Typical distributions in the number of Cherenkov photons recorded by
individual PMTs in 300 MeV electron events and 500 MeV muon events.}
\label{PMT}
\end{center}
\end{figure}

Therefore, we can characterize their classes by the mean vector
 ${\bar{Q_j}(\omega_i, E_0, {\bf r}_0, {\bf \theta}_0)}$ and the
covariance matrix \\
$\Sigma_Q(\omega_i, E_0, {\bf r}_0, {\bf \theta}_0 ) = cov (Q_j, Q_m)$
 and treat the joint distributions of
the contributions $Q_j$ as multidimensional normal distributions:
$$
p({\bf Q}; \omega_i,E_0,\overrightarrow{r_0},
\overrightarrow{\theta_0})
= \,(2\pi)^{-N/2}\cdot\left(det\, \Sigma_{\omega_i}^{-1} \right )^{-1/2}
 \cdot exp\left\{-\left({\bf Q} - \bar{\bf Q}^{\omega_i} \right )^T
\Sigma_{\omega_i}^{-1} \left({\bf Q}-\bar{\bf Q}^{\omega_i}
\right ) \right \} \;  \; \;\;\;\; (1)
$$

Here $\bar{\bf Q}^{\omega_i} \, = \, 
\bar{\bf Q}
(\omega_i,E_0,\overrightarrow{r_0},\overrightarrow{\theta_0})$,
$\Sigma_{\omega_i} \, = \, \Sigma_Q(\omega_i,E_0,\overrightarrow{r_0},
\overrightarrow{\theta_0})$.

    Yet, some event sample is required for calculating $\bar{{\bf Q}}$ and $\Sigma_Q$
     in this case, but its size can be several
orders of magnitude smaller than that in the general case, about some tens of events.
 Within our approach, it is assumed that
the mean vector of an image, $\bar{{\bf Q}}$, and the vector of fluctuations,
 $\delta {\bf Q}$, can be calculated with the aid
of approximations of the mean values and fluctuations of the angular distributions
 of Cherenkov light. In electron (muon)
events, typical values of the correlation coefficients change from 0.6 (0.8)
 for neighboring cells to 0.1 (0.1) for distant
ones. As a result, the covariance matrix proves to be close to a diagonal
 matrix (more precisely, to a sparse matrix): from
the outset, we treat the correlation coefficients of about 0.1 as vanishing
 ones, whereupon each photomultiplier tube has only
four neighbors having nonzero correlation coefficients. We performed test
 calculations with such a covariance matrix, as well
as with a diagonal matrix. The results demonstrated an insignificant
 difference in the quality of event-type identification
for these two forms of matrices. Therefore, we can neglect any
 correlations between $Q_j$ to simplify the form of the
probability density corresponding to a multidimensional normal distribution.
$$
p(\bar{\bf Q}; \omega_i,E_0,{\bf r_0},{\bf \theta_0}) \,
= \, (2 \pi)^{-N/2} \cdot \left (\prod \limits_{j=1}^N \delta Q_j^
{\omega_i} \right)^{1/2} 
\cdot exp \left \{ - \sum \limits_{j=1}^N \frac{\left ( Q_j - Q_j^{\omega_i}
 \right )^2}{\delta Q_j^{\omega_i}} \right \} \; , \; \; \; \; (2)
$$

    For given event geometry specified by the injection point ${\bf r}_0$
and particle direction ${\bf \theta}_0$ and
given energy $E_0$, we calculated the image patterns for the mean value,
$Q_j^{\omega_i} = Q_j^{e,\mu}(E_0, {\bf r}_0, {\bf \theta}_0)$, and for fluctuations, \\
$\delta Q_j^{\omega_i} = \delta Q_j^{e,\mu}(E_0 , {\bf r}_0 , {\bf \theta}_0)$, by the formulae
%
%
$$
Q_j^{e,\mu}(E_0,{\bf r_0},{\bf \theta_0}) \, 
= \, \sum
\limits_{k=1}^n \frac{S}{D_{j,k}^2} \cdot cos \chi_{j,k} \cdot
exp \left(-\frac{D_{j,k}}{\lambda_{abs}} \right) \cdot
F_k^{e,\mu}(\theta_{j,k}) \; ,
\; \; \; \; \; \; \; \; \; (3)
$$

$$
\delta Q_j^{e,\mu}(E_0,{\bf r_0},{\bf \theta_0})
 \, = \, \sum
\limits_{k=1}^n \left [\frac{S}{D{j,k}^2} \cdot cos \chi_{j,k} \cdot
exp \left(-\frac{D_{j,k}}{\lambda_{abs}} \right) \right ]^2 
\cdot \left [ F_k^{e,\mu}(\theta_{j,k}) \cdot
\delta_k^{e,\mu}(\theta_{j,k}) \right ]^2 \; ,
\; \; \;  (4)
\vspace{0.2cm}
$$

where k is the segment index; n is the number of segments in the track (shower); S is the area of the circle
representing a photomultiplier; $D_{j,k}$ is the distance from the segment center to the photomultiplier
center; $cos \chi_{j,k}$ is the cosine of the angle between the vector ${\bf D}_{j,k}$ and the photomultiplier
axis, which is normal to the tank surface; $\theta_{j,k}$ is
the emission angle [that is, the angle between the
track (shower) axis and the vector ${\bf D}_{j,k}$;
and $\lambda_abs$ is the light absorption length in water.
This formula for fluctuations suggests the absence
of correlations between the contributions of individual segments,
this being close to the actual situation since, in the samples
of simulated electron events, the absolute values
of the correlation coefficients do not exceed 0.4 for neighboring segments and are about 0.1
for more distant segments.
In the muon events, typical absolute values of the correlation coefficients are still lower:
they are about 0.1 even
for neighboring segments; this may not be so only for the last segments of the track:
radiation from them is 100 times less intense than from
other segments, but it is correlated because of muon decay.
                                                          
    Once the typical features of the classes have been determined, we can formulate a statistical test for
identifying the event type. We use the Bayes decision rule,
which minimizes the decision error \cite{Fukunaga}. Under
the assumption that the a priori probabilities for the
electron and muon arrival are equal to each other (this   
corresponds approximately to the expected relation between
the fluxes of these events), the ratio of the 
conditional probabilities for the electron and muon arrival
in the case of recording the image ${\bf Q}$ can be
represented in the form
$$
r \, = \, \frac{P \left (e/{\bf Q} \right )}
{P \left (\mu/{\bf Q} \right )} \, 
= \,
\frac{p \left ({\bf Q}/e \right )}
{p \left ({Q}/\mu \right )} \, 
= \,
\frac{\left (\prod \limits_{j=1}^N \delta Q_j^{e} \right )^{1/2}}
{\left (\prod \limits_{j=1}^N \delta Q_j^{\mu} \right )^{1/2}} \cdot
\frac{exp \left \{ - \sum \limits_{j=1}^N \left ( Q_j - Q_j^{e} \right )^2/
\delta Q_j^{e} \right \}}
{exp \left \{ - \sum \limits_{j=1}^N \left ( Q_j - Q_j^{\mu} \right )^2/
\delta Q_j^{\mu} \right \}} \; ,
\; \; \; (5)
$$
where $p \left ({\bf Q}/e \right )$ and
$p \left ({\bf Q}/\mu \right )$ come from Eq.(2).
%
%

     The simplest criterion used to identify the
    event type is
$$
q \, = \, 2 ln \, r \, = \, q_{\mu} \, - \, q_{el} \, + \, C \;
 , \; \; \; \; \; \; \; \; \;(6)
$$
$$
q_{\mu} \, = \, \sum \limits_{j=1}^N \left ( Q_j^{\, exp} \, - \, Q_j^{\, \mu}
\right )^2 / \delta Q_j^{\, \mu} \; ,
$$
$$
q_{el} \, = \, \sum \limits_{j=1}^N \left ( Q_j^{\, exp} \, - \, Q_j^{\, e}
\right )^2 / \delta Q_j^{\, e}  \; ,
$$
$$
C \, = \, \frac{1}{2} \, ln \, \frac{\prod \limits_{j=1}^N \delta Q_j^{e}}
{\prod \limits_{j=1}^N \delta Q_j^{\mu}}  \; ,
$$
%
 where $Q_j^{expt}$ is the light contribution to the $j$th photomultiplier in the experimental image being
considered. The event is assumed to be of the $e$ or $\mu$ type
if $q > 0$ or $q < 0$, respectively; in the case of   
$q = 0$, the event is rejected.                            
                                                         
    A somewhat more general form of the criterion can reduce the decision error in event-type identification.
It can be taken in the form              
$$
q = q_{\mu} - A q_e + B,  \;\;\;\;\;   (7)
$$
where A and B are tuned to minimize the identification errors. The application of this statistical test is
similar to the application of the test used above.

\begin{figure}
\begin{center}
\rotatebox{0}{ \resizebox{0.45\textwidth}{!}{\includegraphics{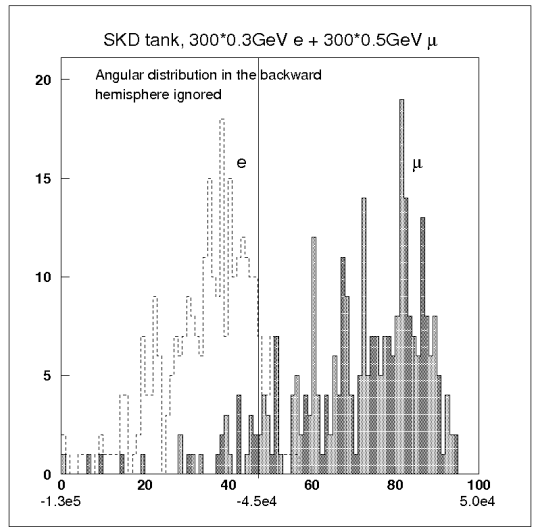}}}
\caption{Typical distributions in $q_{\mu} - q_e$ criterion applied to artificial
300 MeV electron events and 500 MeV muon events. Cherenkov light angular distribution
approximated in the range $0^o - 90^o$.}
\label{typefw}
\end{center}
\end{figure}
                                                         
    Figure 4 shows typical distributions of $q_{\mu} - q_e$
for simulated electron and muon events characterized by
energies of 300 MeV and 500 MeV, respectively. The optimum boundary between the electron and muon
samples for the first statistical test is indicated by the vertical line (that is, C is determined in this case
by minimizing the identification error rather than by using the respective formula). The error associated
with misidentifying an electron as a muon, as well as the error associated with misidentifying a muon
as an electron, is about 10\%. An analysis of events in the long tail of the muon distribution revealed that
$q_{\mu} - q_e$ values for muons that are similar to those
in $e$ events are due to decay electrons, which generate
an isotropic mean distribution of light. The error of event-type identification can be reduced by using the
light angular distribution in the range $0^o - 180^o$ rather
than in the range $0^o - 90^o$ to calculate the pattern
image. Figure 5 shows the $q_{\mu} - q_e$ distributions
for the same event samples as in Fig. 4, but, in calculating
the pattern image, the mean angular distributions and
fluctuations of light emitted into the backward hemisphere are
assumed to be constant and equal to their values at $\theta = 90^o$ rather than ignored,
 as was done in the first case.
It can be seen that the distributions of the two event classes are separated much better in this case: after
 optimization,
the error associated with misidentifying an electron as a muon becomes as small as a few tenths of a percent,
 while the error
associated with misidentifying a muon as an electron proves to be 2 to 3\%.                                  

\begin{figure}
\begin{center}
\rotatebox{0}{ \resizebox{0.45\textwidth}{!}{\includegraphics{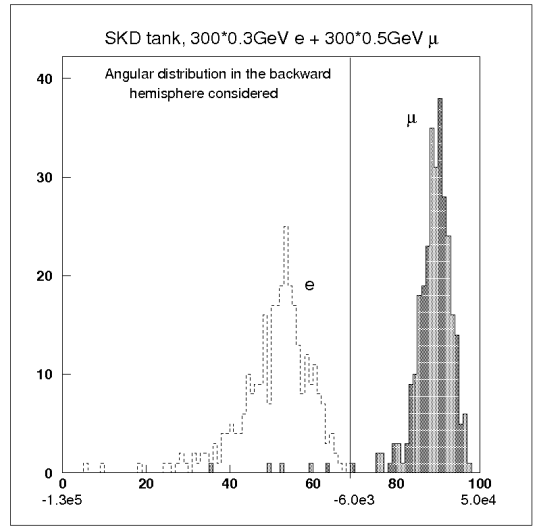}}}
\caption{Typical distributions in $q_{\mu} - q_e$ criterion applied to artificial
300 MeV electron events and 500 MeV muon events. Cherenkov light angular distribution
approximated in the range $0^o - 180^o$.}
\label{typefwbw}
\end{center}
\end{figure}

\begin{figure}
\begin{center}
\rotatebox{0}
{ \resizebox{0.6\textwidth}{!}{\includegraphics{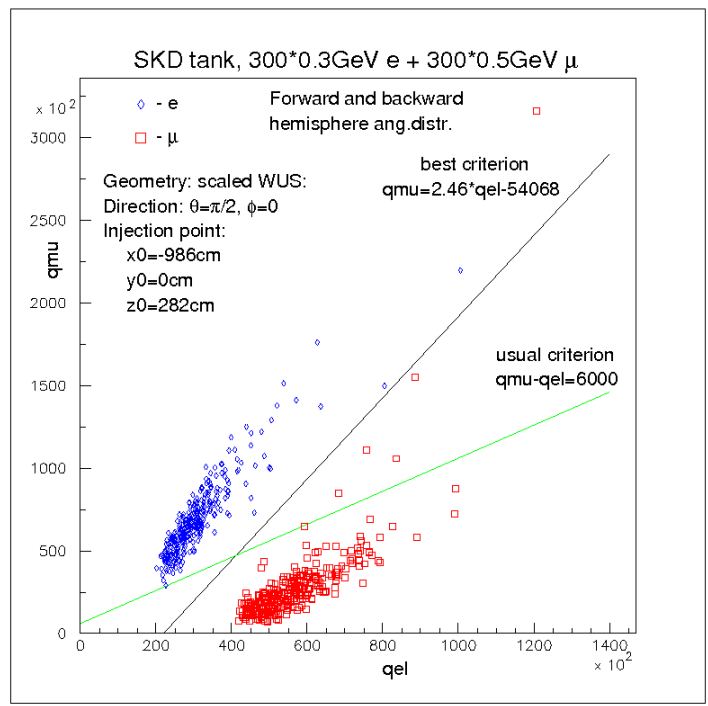}}}
\caption{Correlation plot $q_{\mu}$ --- $q_e$ for artificial
300 MeV electron events (diamonds) and 500 MeV muon events (squares).
Cherenkov light angular distribution approximated in the range $0^o - 180^o$.
The green line represents a simple criterion shown in Fig~\ref{typefwbw},
while the black one shows the best criterion.}
\label{typeplot}
\end{center}
\end{figure}

    The application of the more general criterion permits us to improve even this result
(Fig.~\ref{typeplot}).
The criterion (7) optimized
throughout the sub-GeV range, which embraces $e$ events at 300 MeV ($\mu$ events at 500 MeV),
 $e$ events at 500 MeV ($\mu$
events at 700 MeV), and $e$ events at 700 MeV ($\mu$ events at 900 MeV) makes
it possible to reduce both identification errors to fractions of a percent.

    Thus, the upper limits obtained for the errors in
event-type identification are in reasonable agreement 
with the errors estimated by the SuperKamiokande
Collaboration for sub-GeV energies \cite{Kasuga3}: 0.5\% and 1.0\% for,
respectively, electron and muon events.
                                                        
    The above results illustrate the application of the statistical test to events characterized by the lowest
of the energies considered here, this case being the most complicated for classification. The difference in
the Cherenkov images for muons and electrons becomes more pronounced with increasing energy, this facilitating
 event-type identification.

\section{Reconstruction of Event Geometry}

\subsection{TDC procedure}

The SK analysis introduces a ``TDC procedure'' to determine the vertex
position.  The principle of the TDC procedure is to find the position
where the time residuals, $t_i$, for the PMTs being fitted are
minimized.  The time residual $t_i$ of the i-th PMT is defined as
$$
t_i \, = \, t_i^0 \, - \, \frac{n}{c} \times
\sqrt{(x-x_i)^2 \, + \, (y-y_i)^2 \, + \, (z-z_i)^2}
\; , \; \; \; (8)      
$$
where $t_i^0$ is the hit time of the $i$-th PMT, $(x_i,y_i,z_i)$ is
the position of the $i$-th PMT, $(x,y,z)$ is the effective emitting
point position and $c/n$ is the velocity of the Cherenkov light in
water.  That all the light is emitted from the same {\it effective}
point in space and comes to $j$-th PMT {\it exactly} at the moment
corresponding to the mean time $\bar{t_j}$ of the $j$-th PMT Cherenkov
pulse is not really true. A simple equation for time residual used in
a $\chi^2$-like sum (the system of linear equations for the effective
point coordinates $(x^*,y^*,z^*)$ is overdefined!) can give effective
point estimate after the sum's minimization (with respect to
$(x^*,y^*,z^*)$) even if the original assumption is not valid. The
effective point thus deduced does not coincide with the center-of mass
of the light emitting system ($e$-shower or $\mu$-track) because of
specific mechanism of Cherenkov pulse formation and will usually
differ from the event starting (injection) point.

The TDC procedure based only on $\bar{t_j}$ cannot estimate the event
direction because to define a direction one needs at least two
points. Direction estimates could be obtained as a result of Cherenkov
pulse shape analysis for each sufficiently illuminated optical module
if the PMT and electronics are fast enough for such analysis.

\begin{figure*}
\begin{center}
\rotatebox{90}{\resizebox{0.6\textwidth}{!}{\includegraphics{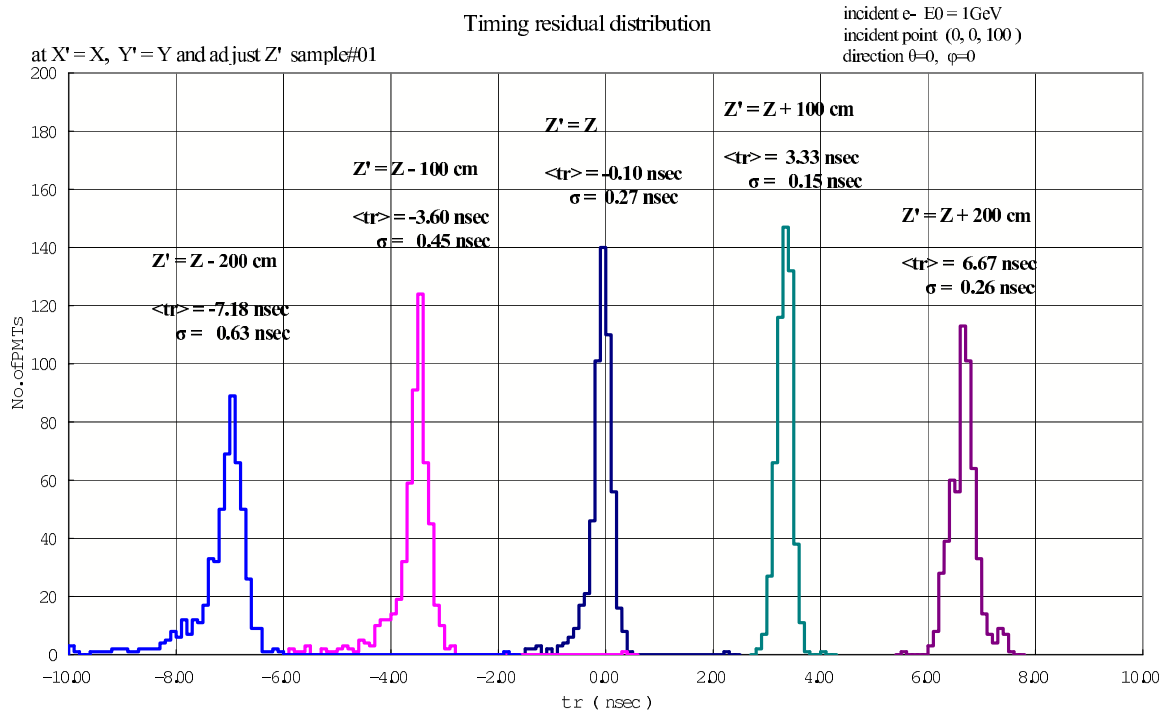}}}
\caption{\label{fig:12}  One example of the time residual distributions for 
1 GeV electron primary shower, assuming different vertex positions.}
\end{center}
\end{figure*}

Sakai shows the time residual distribution of typical event (a
1~GeV/c, electron) which is distributed over 50 nanoseconds (Sakai, p.38
 \cite{Sakai}), assuming a point-like source. We simulate
the Cherenkov light in the cascade shower using GEANT~3.21 and the
tools we have developed.  In 
Figure\ref{fig:12}
we give one example for the
time residual distribution for a 1~GeV primary electron based on
Eq.(8) with the use of the detailed simulation of the cascade shower.
In our calculation, we simulate shower particles and the accompanying
Cherenkov light due to shower particle concerned. Then, we know the
starting point of the primary electron.
Shifting the starting point from the real point to as range of
artificial ones, we can obtain the time residual distribution for each
position, and examples are given in 
Figure\ref{fig:12}.\\

Among the five
different starting points, which includes the true one, the smallest
standard deviation is obtained in the case of $ Z'=Z+100 $~(cm), where
$Z'$ and $Z$ denote the assumed vertex point and the real vertex
point, respectively. Thus, the apparently most probable vertex point
is not real one, but is offset by 100~cm from the real one.  Of
course, this is only one example and not average behavior.  However we
examined many individual cases and confirm that this is usual
character which should hold even for the average behavior.

The comparison of our simulations with the experimental data from the
SK experiment (Sakai) reveals a large difference.  The width of our
time residuals distribution is within one nanosecond, while the width
for SK is $\sim$50~ns.  The reasons are that we have
not considered the PMT and electronics response functions, and that we
have neglected light scattering.

We calculate the time residuals for electrons of 500~MeV, 1~GeV, 3~GeV
and 5~GeV, assuming that all Cherenkov light comes from certain points
of a shower/track. From these calculations one can see that the
smallest time residuals do not give the vertex position but yield
points shifted from the vertex point along the direction of the
cascade shower, namely, 50 to 100~cm for 500~MeV electrons, 100 to
150~cm for 1~GeV
(Figure\ref{fig:12})
 and 3~GeV electrons, and 150 to 200~cm
for 5~GeV electrons.  Such a tendency is quite understandable if we
consider the size of the shower/track: the effective point should not
be too close to the starting or ending points.  The error of effective
point location by minimizing the width of the distribution amounts to
about 50~cm.

    From the much larger width of the SK time residual distribution it
is clear that in experimental conditions the effective point location
error should be a few times greater because the minimum of the width
as a function of effective point position would be much less
pronounced.

For reasons mentioned above, we conclude that the SK TDC procedure is
not suitable for the determination of an accurate vertex position for
electron events.  The situation for the muon events is essentially the
same but must be worse than in the case of electron events as muon
events have a longer extent than the corresponding electron events.

\subsection{ADC procedure}

\subsubsection{Procedure for Geometry Reconstruction}

    Within the procedure used to reconstruct event geometry, it is assumed that the type of an event and
its energy are known. Thus, one simultaneously seeks only the injection point ${\bf r}_0$ and the direction
 ${\bf \theta}_0$ of primary particle.
                                                        
    At the first step of the procedure, ring-shaped (arc-shaped) structures are sought in the optical image.
     In the case
of a muon event, two ring-shaped or spot-shaped structures are
observed since an electron produced in muon decay also emits
Cherenkov light. We did not consider images from muons of momentum below 500 MeV/c and therefore selected the
 more intense of
the two structures. Two-dimensional 
 (or column-by-column)
scanning in order to find the maximum above some threshold in the number of Cherenkov photons.

    At the second step, the first ${\bf r}_1$ and ${\bf \theta}_1$ approximations to the geometric
     event parameters were
determined by approximating the ring-shaped structure by the following simple conelike model of the event.
The entire amount of light Qtot emitted by a $\mu$ track ($e$ shower) is assumed to originate from
a single point W on the track (shower axis) and to have an angular
distribution of the form 

$$
F(\theta) \, = \, \left \{ 
\begin{array}{cccc}
& 0, &  \theta \, < \bar{\theta} - \Delta \theta   & \\
& a, &  |\theta - \bar{\theta}| \le \Delta \theta  & \\
& 0, &  \theta \, > \bar{\theta} + \Delta \theta  & 
\end{array} \right.,
 a:\; \; 2\pi \, \int \limits_0^{\pi} F(\theta)
 \, sin \theta d\theta \, = \, Q_{tot} \; \; \; \; (9)
$$

    In this case, a zero-order approximation can be arbitrary since a ring-shaped structure is usually quite
distinct, while the conelike model of the Cherenkov light distribution has sharp edges. The approximation
is performed by means of a numerical minimization of the
following function with respect to the variables ${\bf r}$
and ${\bf \theta}$:
$$
G(\mathbf{r},\mathbf{\theta}) \, = \,
\sum \limits_{l=1}^M \frac{\left ( Q_l^{\, expt}(\mathbf{r_0},
\mathbf{\theta_0}) \, - \, Q_l(\mathbf{r},
\mathbf{\theta}) \right )^2}{Q_l^{\, expt}(\mathbf{r_0},
\mathbf{\theta_0})}
\; , \; \; (10)
$$

Here, $l$ is the photomultiplier index within the ring-shaped
structure; M is the number of photomultipliers in the
ring-shaped structure; $Q_l^{expt}$ is the light contribution
to the $l$th photomultiplier tube from the event being
considered; and $Q_l ({\bf r}, {\bf \theta})$ is the estimate of this contribution according to the
calculation within the above cone-like model $F(\theta)$ of the light angular distribution,
$$
Q_l(\mathbf{r},\mathbf{\theta}) \, = \,
 \frac{S}{D_{l,W}^2} \cdot cos \chi_{l,W} \cdot
exp \left(-\frac{D_{l,W}}{\lambda_{abs}} \right) \cdot
F(\theta_{l,W}) \; , \; \; (11)
$$


\begin{figure*}
\vspace{0,5cm}
\hspace*{1.5cm}
\rotatebox{90}{ \resizebox{0.55\textwidth}{!}{\includegraphics{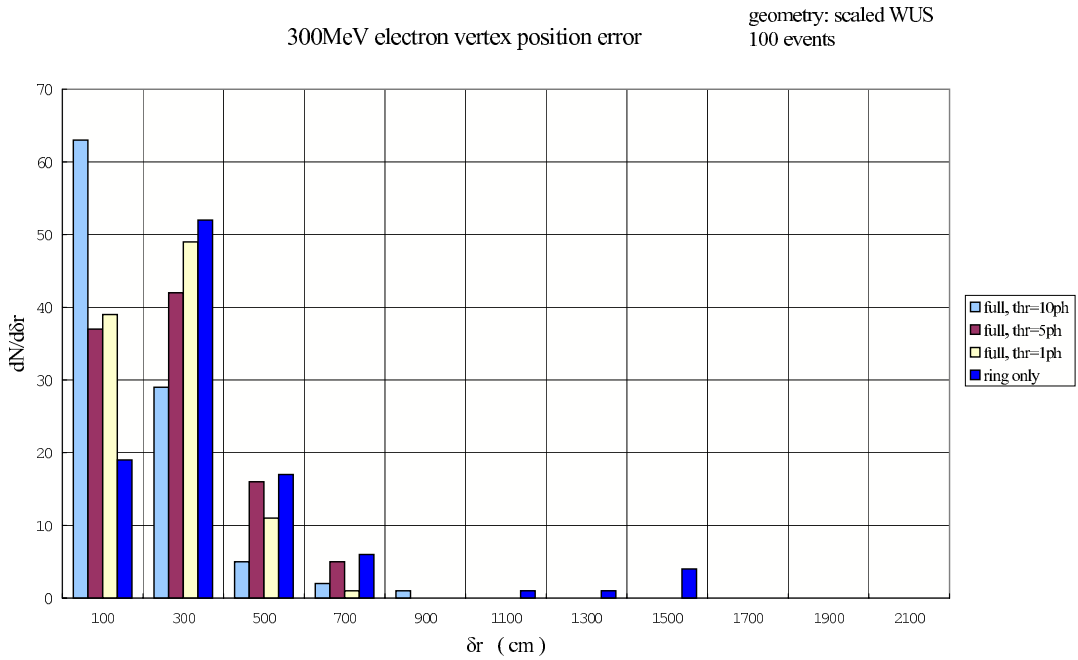}}}
\caption{\label{fig:15}
Error distribution for the vertex position for 300 MeV 
electrons injected at the scaled WUS point. 
The sample volume is 100.}
\end{figure*}

\begin{figure*}
\vspace{0.5cm}
\hspace*{1.5cm}
\rotatebox{90}{\resizebox{0.55\textwidth}{!}{\includegraphics{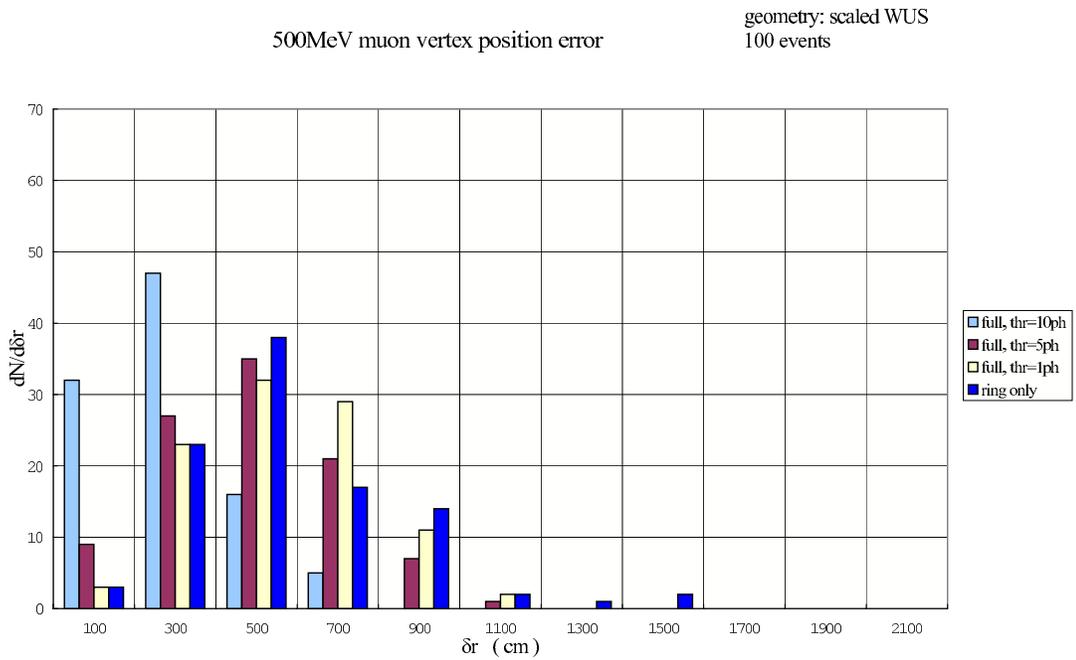}}}
\caption{\label{fig:16} Error distribution for the vertex position for 500 MeV
muons injected at the scaled WUS point. The sample volume is 100.}
\end{figure*}

    The third (last) step of the procedure consists in improving the estimates of the geometric parameters
by approximating the whole image (including only the contributions above some threshold $Q_{thr}$) by the
pattern image calculated for the corresponding event class,
{$Q_j^{e,\mu}(E_0, {\bf r}, {\bf \theta})$, $\delta Q_j^{e,\mu}(E_0, {\bf r}, {\bf \theta})$},
 within a
detailed model of the Cherenkov light angular distribution as was described in the preceding section. The
first approximation obtained at the second step of the procedure is used as the zero-order approximation for
the approximation being considered. Specifically, one performs a numerical minimization of the following
function with respect to ${\bf r}$ and ${\bf \theta}$:

$$
H(\mathbf{r},\mathbf{\theta})  
= \sum \limits_{j: \, Q_j^{\, expt} \ge Q_{thr}}
\frac{\left ( Q_j^{\, expt}(E_0,\mathbf{r_0},
\mathbf{\theta_0}) \, - \, Q_j(E_0,\mathbf{r},
\mathbf{\theta}) \right )^2}
{\delta Q_j(E_0,\mathbf{r},\mathbf{\theta})} \; \; \; (12)
$$


    An optimum threshold for the Cherenkov contribution to photomultiplier,
$Q_{thr,op}$, can be chosen in such a way
as to minimize the uncertainty in determining the geometric parameters
(the geometric resolution accordingly being
maximal in this case).
The optimum threshold grows with increasing primary energy.
This can be used to improve the resolution in an actual
experiment since the energy can be estimated on the basis of the total amount of recorded light.

\subsubsection{Results of our Analysis}
By using the developed technique we analyze
simulated events to determine the error distributions for the vertex
position and for the particle direction.  Here, we examine the error
distributions for 300~MeV electrons and 500~MeV muons, which yield
roughly the same amount of Cherenkov light.

\begin{figure*}
\hspace*{1.0cm}
\rotatebox{90}{\resizebox{0.58\textwidth}{!}{\includegraphics{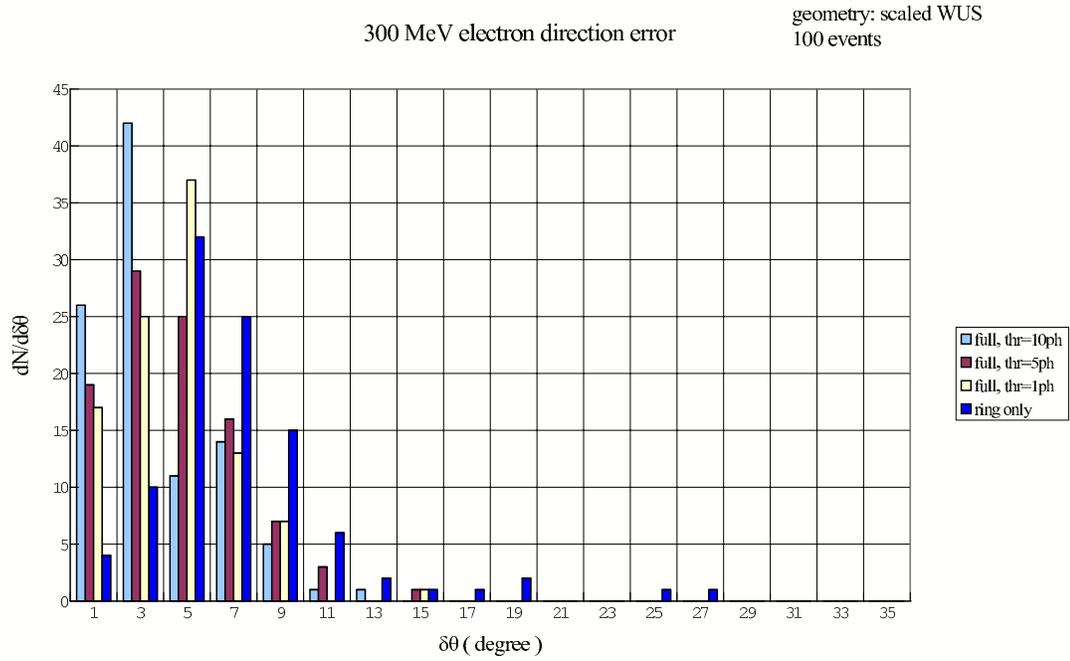}}}
\caption{\label{fig:17} Error distribution for the direction for 300 MeV
 electrons injected at the scaled WUS point. The sample volume is 100.}
\end{figure*}

\begin{figure*}
\vspace{0.5cm}
\hspace*{1.0cm}
\rotatebox{90}{\resizebox{0.58\textwidth}{!}{
\includegraphics{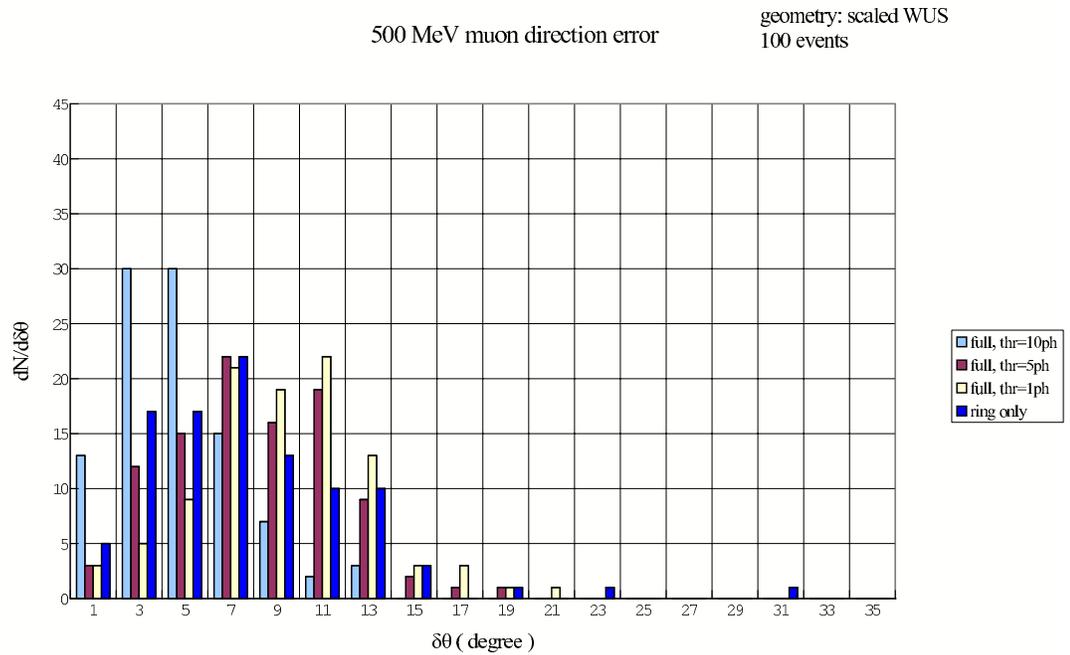}}}
\caption{\label{fig:18} Error distribution for the direction for
 500 MeV muons injected at the scaled WUS point. The sample volume is 100.}
\end{figure*}

In Figure~\ref{fig:15}, we give the error distribution for the vertex position
for 300~MeV electrons for different Cherenkov threshold quantities.
``Ring only'' denotes that only the information from PMTs whose
Cherenkov photons contribute to the Cherenkov ring 
are used for the estimation of the error.  ``Full proc., thr=1ph'' 
denotes that information from ``ring only'' PMTs and also those
exceeding 1 Cherenkov photon are utilized for the estimation on the error. 
For ``full proc., thr=5'' and ``full proc., thr=10'', this latter threshold
is raised to 5 and 10 photons, respectively.

In Figure~\ref{fig:16} the error distribution for the vertex position for 500~MeV 
muons are given. It is clear
that a wider error distribution is obtained for a ``ring only''
analysis.  A narrower error distribution results from the ``full proc,
thr=10 ph'' algorithm. This is the same as in the case of the
electron.  However, muons generally have wider error distributions
than electrons: the mean error for 500~MeV muons for the vertex
determination in the full analysis is 2.9~m while for 300 MeV electrons it amounts to 2 m.

In Figure~\ref{fig:17}, the error distribution for the direction of the
300~MeV electron is given. As expected, ``ring only'' gives the largest
uncertainty distribution, while ``full proc., thr=10ph'' has the
narrowest error distribution, with a mean error of about 3.7$^\circ$.
In Figure~\ref{fig:18},
 we give the corresponding distributions for muons. The
same trend is seen as for electrons, though the muons have a wider
uncertainty distribution.  The mean direction uncertainty in the best case is 4.9$^\circ$.

\begin{figure*}
\hspace*{1.5cm}
\rotatebox{90}{\resizebox{0.55\textwidth}{!}{\includegraphics{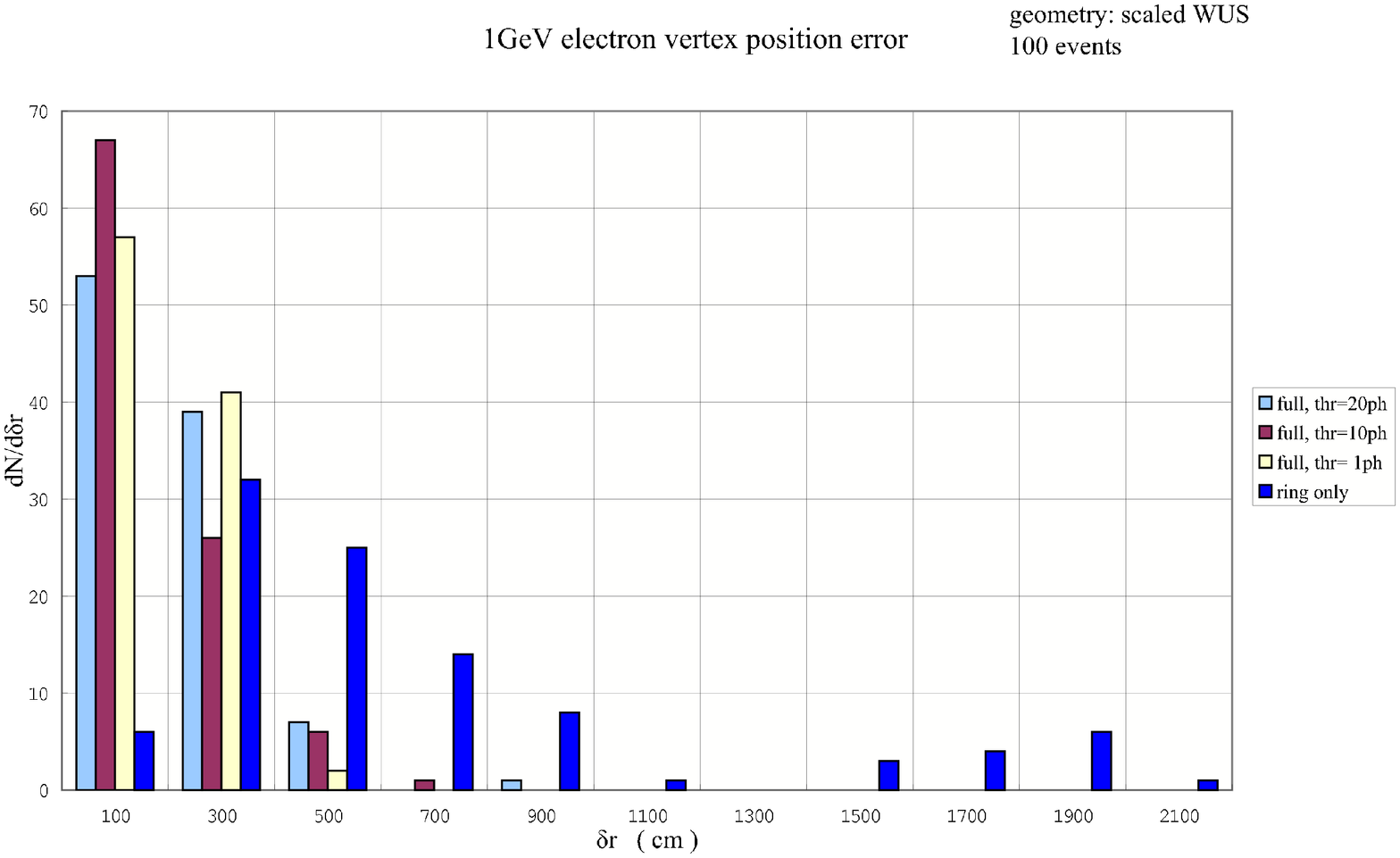}}}
\caption{\label{fig:19} Error distribution for the vertex position for 1 GeV
 electrons injected at the scaled WUS point. The sample volume is 100.}
\end{figure*}

\begin{figure*}
\hspace*{1.5cm}
\vspace{0.5cm}
\rotatebox{90}{\resizebox{0.55\textwidth}{!}{\includegraphics{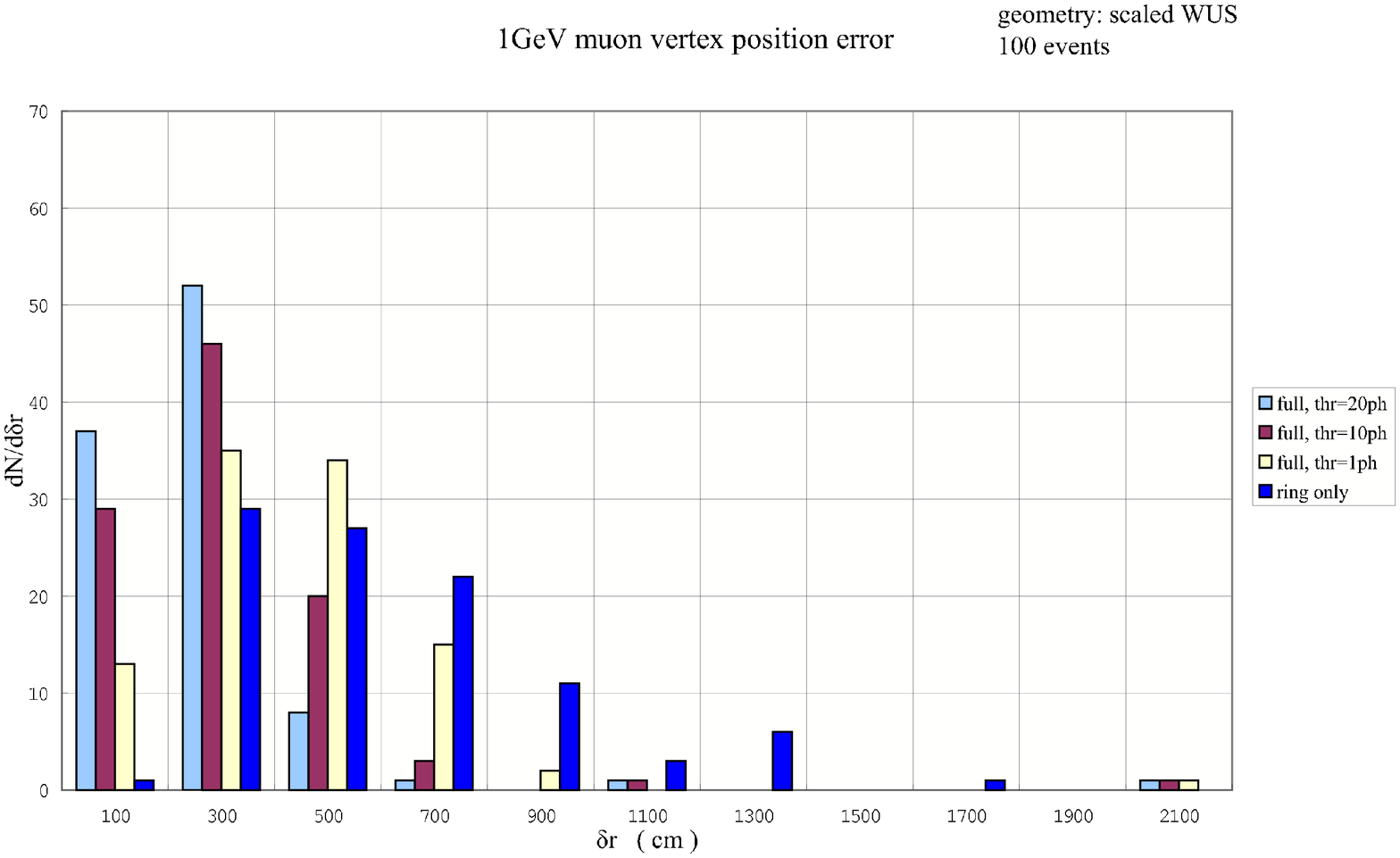}}}
\caption{\label{fig:20} Error distribution for the vertex position 1 GeV 
 muons injected at the scaled WUS point. The sample volume is 100.}
\end{figure*}

\begin{figure*}
\hspace*{1.5cm}
\rotatebox{90}{\resizebox{0.58\textwidth}{!}{\includegraphics{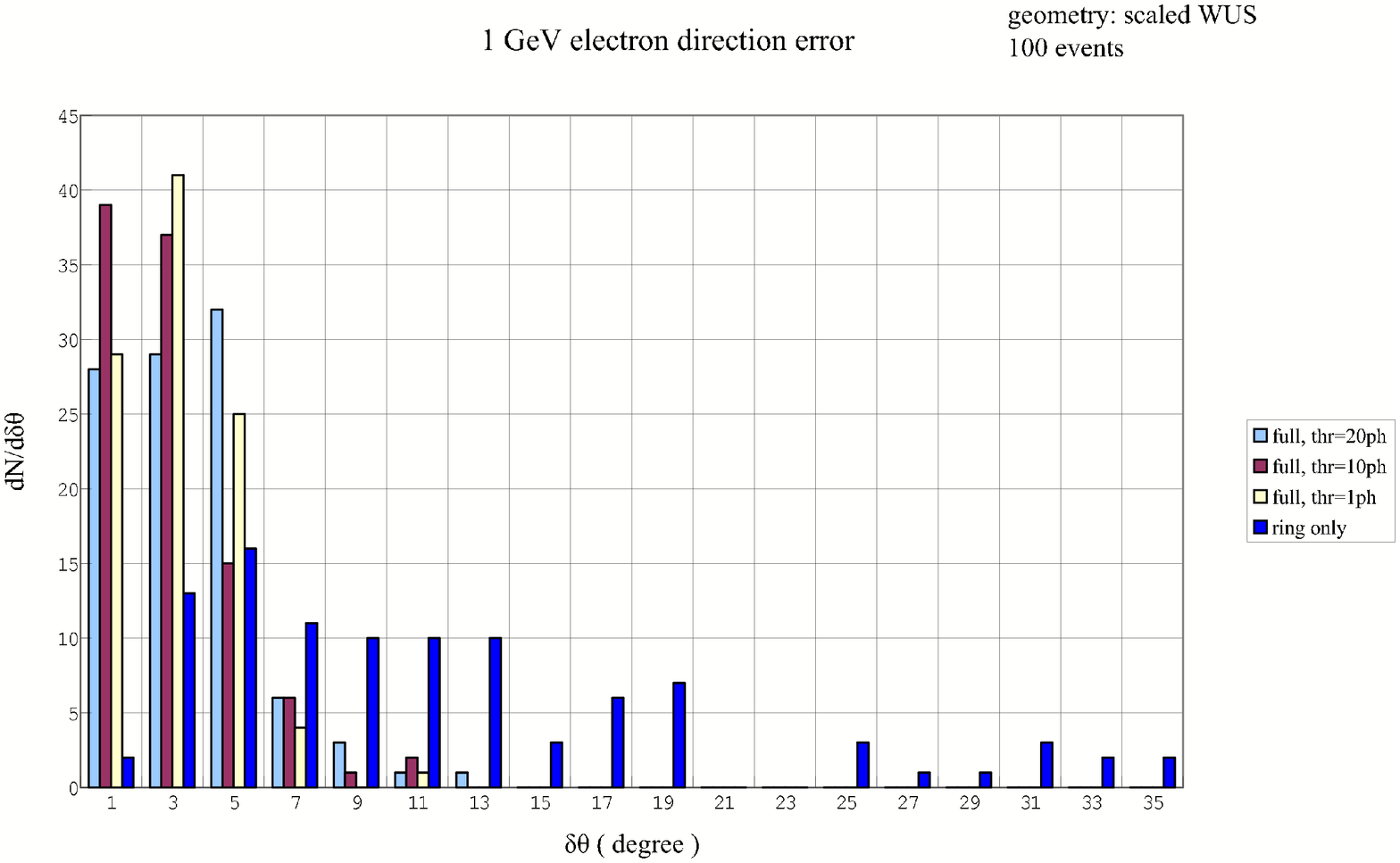}}}
\caption{\label{fig:21} Error distribution for the direction for 1 GeV
 electrons injected at the scaled WUS point. The sample volume is 100.}
\end{figure*}

\begin{figure*}
\vspace{0.5cm}
\hspace*{1.5cm}
\rotatebox{90}{\resizebox{0.58\textwidth}{!}{\includegraphics{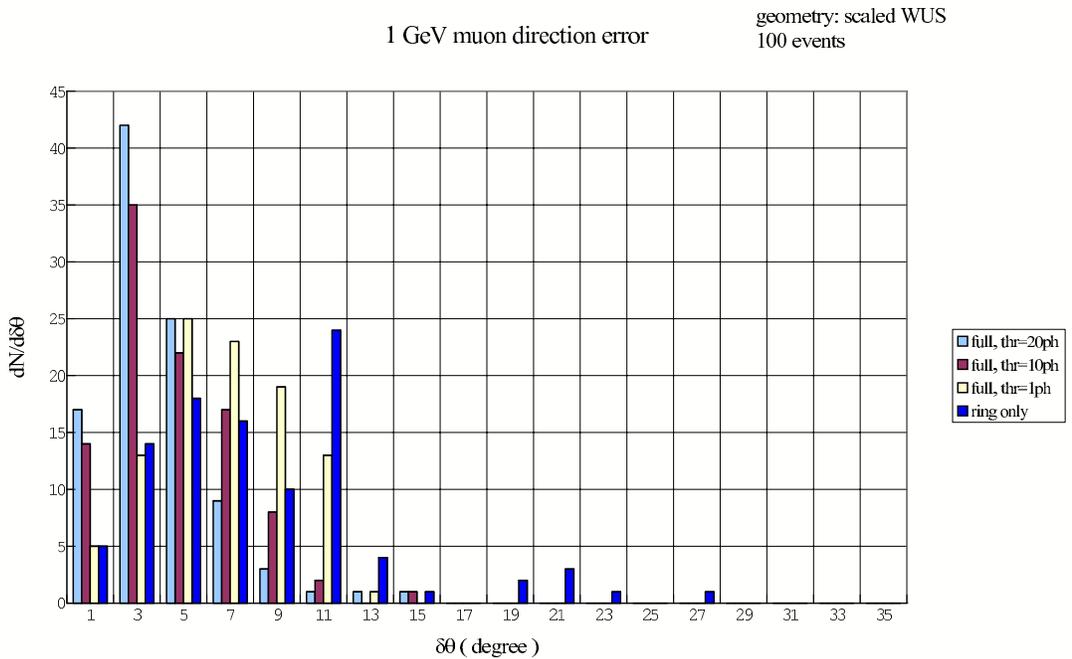}}}
\caption{\label{fig:22} Error distribution for the direction 1 GeV 
 muons injected at the scaled WUS point. The sample volume is 100.}
\end{figure*}

Now, we compare 1~GeV electrons with 1~GeV muons, both of which yield
roughly the same quantity of Cherenkov light.  In
Figure~\ref{fig:19}, we give the error distribution for the vertex
position for 1~GeV electrons in the case of ``full proc,thr=10''. The
mean error is 1.9~m.  In Figure~\ref{fig:20}, the corresponding
quantities for the muon are plotted.  The average error for the vertex
position is 3.2~m.  Again, the error of the vertex point for muons is
larger than for electrons.

In Figure~\ref{fig:21}, the error distribution for the direction for
1~GeV electron is shown. The average direction error is 3.0$^\circ$
for ``full proc,thr=10''.  The corresponding quantities for the muon
are plotted in Figure~\ref{fig:22}, where the mean error is
5.3$^\circ$. Once again, the directional error for muons is larger
than that for electrons.

\begin{center}
\begin{table*}
\caption{\label{Table:3} 
Mean and standard deviation of the error in the vertex position and the
direction due to primary electrons and primary muons. These are given
 for different criteria for the Cherenkov threshold.  Ring proc. 
 denotes errors estimated using the Cherenkov 
 ring only. [1], [5], [10], [20] denote errors estimated by the 
 combination of [ring proc.] with a Cherenkov photon threshold 
 of 1, 5, 10 and 20photons, respectively.  Alm denotes the mean 
 direction error in degrees. Als denotes the standard deviation 
 for the corresponding mean values. Rm denotes the mean position
  error in metres. Rs denotes the standard deviation for the 
  corresponding mean values.
 }
\vspace{5mm}
\hspace*{4cm}
\begin{tabular}{|c|c|c|c|c|c|c|}
\hline
&&\multicolumn{5}{c|}{$Q_{thr}$, threshold}\\
\cline{3-7}
&&Ring proc.&[1]&[5]&[10]&[20]\\
\hline
         & Alm (deg.) &  7.2   &    4.4   &   4.6   &   3.7   &        \\
300 MeV  & Als (deg.) &  4.2   &    2.4   &   2.7   &   2.5   &        \\
electron &  Rm (m)   & 3.88   &    2.50  &   2.74  &   1.98  &        \\
         & Rs (m)    & 2.91   &    1.26  &   1.60  &   1.49  &        \\
\hline

         & Alm (deg.) &  7.8   &    9.3   &   8.0   &   4.9   &        \\
500 MeV  & Als (deg.) &  4.7   &    3.7   &   3.6   &   2.7   &        \\
muon     &  Rm (m)   &  5.71  &    5.69  &   4.92  &   2.94  &        \\
         &  Rs (m)   &  2.57  &    2.01  &   2.17  &   1.54  &        \\
\hline

         & Alm (deg.) & 11.7   &    3.1   &         &   3.0   &    3.7  \\
 1 GeV   & Als (deg.) & 14.4   &    3.5   &         &   3.7   &    4.4  \\
electron &   Rm (m)  &  6.49  &    1.99  &         &   1.86  &    2.17 \\
         &  Rs (m)   &  8.21  &    2.19  &         &   2.24  &    2.55 \\
\hline

         & Alm (deg.) & 8.8    &    7.3   &         &   5.3   &    4.9  \\
1 GeV    &Als (deg.)  &11.3    &   11.1   &         &  10.0   &   10.0  \\
muon     & Rm (m)    & 5.98   &    4.25  &         &   3.15  &    2.70 \\
         & Rs (m)    & 6.76   &    4.92  &         &   3.99  &    3.62 \\
\hline

         & Alm (deg.) &21.3    &          &         &   2.0   &    1.9   \\
5 GeV    &Als (deg.)  &11.0    &          &         &   5.1   &    4.4   \\
electron &  Rm (m)   &15.07   &          &         &   1.43  &    1.32  \\
         & Rs (m)    & 6.65   &          &         &   2.76  &    2.66  \\
\hline

         & Alm (deg.) & 8.5    &          &         &   4.3   &    3.8   \\
5 GeV    &Als (deg.)  & 6.7    &          &         &   4.0   &    3.8   \\
muon     & Rm (m)    & 4.68   &          &         &   2.89  &    2.45  \\
         & Rs (m)    & 2.52   &          &         &   2.55  &    2.45  \\

\hline
\end{tabular}
\end{table*}
\end{center}

In Table~3, we summarize the error distributions for the vertex points
and the direction for both electrons and muons.  Both mean errors and
root mean square errors are given.  Errors are also given for the
different criteria, namely, different Cherenkov light threshold. 
 From Figures~\ref{fig:15} to \ref{fig:18} and Figures~\ref{fig:19}
to \ref{fig:22} and Table~3, it should be noticed the following:

\begin{enumerate}

\item Of the different criteria considered, the ``ring only''
procedure results in the largest error.  The reasons are as follows:
The concept of the ring structure is essentially fuzzy, both in our
procedure and the SK procedure, and information from ring structure is
only part of the total information available for the pattern
recognition.  It is, therefore, natural that the vertex position and
directional errors are largest in the ``ring only'' analysis.  The
standard SK analysis uses ring structure only, and their errors are
amplified by that fact that the analysis ignores fluctuation effects.

\item Muons events have larger uncertainties than electron events.
For both electrons and muons, the sources for the Cherenkov light are
not point-like and have some extent in both cases. Significant errors
come from the point-like approximation for electron events.

\item The optimal Cherenkov threshold for the third step of geometry 
reconstruction procedure depends on the primary energy of the particle 
concerned. For energies less than 1 GeV, third step with 10 photon 
threshold gives the best results for both  muons and electrons 
among the alternatives considered. For 5 GeV electrons and muons,
 20 photon threshold seems to be optimal for the third step.
 
\item The fact that the uncertainties for the determination of the
vertex point and direction are rather large comes from the effect of
fluctuations, namely the nature of the stochastic process concerned
(an electron cascade shower or sequence of muon interactions with the
medium).  The utility of model developed in this paper, the moving
point approximation model, is guaranteed, because it gives mean values
and relative fluctuations precisely and takes all necessary
geometrical considerations into account correctly.  Even if additional
errors exist, they should be negligible compared to the uncertainty
caused by fluctuations.  The rather large errors for the vertex point
and the direction obtained by our model could not be reduced substantially,
reflecting the essential nature of the physical
processes concerned.

\item The SK analyses, according to all published accounts, completely
neglect fluctuations and also use point-like approximations for the
electron cascade.  Moreover, they neglect the scattering
effects on muon track geometry. As we showed earlier, their simple
approximations distort the mean values in certain parameter
domains. The most probable reason for their low error estimates is the
fact that they completely neglect fluctuations in the event
development.  Our results contradict clearly the fine positional
resolution of 23 to 56\,cm claimed by for the SK analysis
(Kibayashi, p.73\cite{Kibayashi}).

\item It should be noticed that errors derived by us are lower limits.
As already mentioned, we do not consider the production of
photoelectrons in PMTs, and only consider direct Cherenkov photons in our
discrimination procedures neglecting the diffusion of
Cherenkov photons. If we include these factors in our procedure, then,
the actual errors for the vertex position and the direction should be
larger than that given here.

\end{enumerate}

\section{Summary and Conclusion}

\noindent(1) Type definition procedure 

SK procedure for event type definition is based on oversimplified models of events
and is unlikely to give the type definition errors declared by SK. Our procedure,
based on much more accurate event models, is potentially capable of enabling the error
of less than 1\% in type definition. \\

\noindent(2) The SK TDC procedure 

The TDC procedure assumes that the Cherenkov light originates from a
point, and thus does not determine the vertex position accurately,
because the sources for the Cherenkov light have a non-negligible
extent.  In order to utilize the TDC meaningfully, we should take into
account the extent of the source for the Cherenkov light in space and
time. Further, ideally we should utilize not only arrival time of the
Cherenkov light but also shape of the pulse in the PMTs.\\

\noindent(3) Errors for the vertex point and the direction

Our
estimation (Figures~15 to 18 and Figures~19 to 22 and Table~3) 
shows non-negligible and
inevitable errors for the vertex position and direction.  It should
be, particularly, noticed that the fluctuations in error in both
vertex position and direction are rather big.

 Kibayashi (p.73\cite{Kibayashi}) concludes that the uncertainty for
vertex point is from 23~cm to 56~cm and the uncertainty for the
direction from 0.9$^\circ$ to 3.0$^\circ$ using both the estimator for
particle identification and the TDC adopted by the SK.  As we have
demonstrated, these appear to severely underestimate the error
distributions, which is too far from reality.
\\
\noindent (4) In the present paper, we do not take into consideration photoelectrons
produced by the Cherenkov light in the PMT, and we neglect scattering
of the Cherenkov light. Therefore, our results on
discrimination of electron events from muon events, only yield 
lower
limits to the realistically achievable experimental 
errors. \\


   On the basis of a statistical simulation of electron and muon events in a water Cherenkov telescope
close in parameters to the SuperKamiokande telescope, we have constructed realistic models of actual
events. Relying on these models, we have developed algorithms for event type identification and event geometry
reconstruction. Although we did not aim at developing elaborate procedures for this telescope
precisely, our calculations allowed us to obtain upper limits on the resolutions in energy, event type,
and geometric parameters for telescopes of this class.
The observed discrepancies --- first and foremost, the fact that the geometric resolution obtained by the
SuperKamiokande Collaboration is higher than our estimates --- call for future investigations.

Construction of optimal experimental algorithm for SK data is beyond the scope of this paper and is better
undertaken by those with a detailed understanding of the specifics of the detector.

A part of the results of the present paper are found in \cite{Anokhina}.

\section{Acknowledgement}


One of the authors (V.G.) should like to thank Prof.\ M.\ Higuchi,
Tohoku Gakuin University. Without his invitation, V.G.\ could not join
in this work.  Authors would like to be very grateful for the
remarkable improvement of the manuscript to Dr. Philip Edwards.

%

\section*{References}
\begin{thebibliography}{}

\bibitem{Kasuga1} Kasuga {\it et al}, Phys.Lett.{\bf B374},(1996)238 

\bibitem{Fukuda1} Ashie {\it et al},Phys.Rev.{\bf D71},(2005)112005

\bibitem{Olga}
 Ryazhskaya O.G., JETP.Lett.{\bf 60},(1994)609\par
 Ryazhskaya O.G., JETP.Lett.{\bf 61},(1995)229\par
 Khalchukov F.F{\it et al}, Nuovo Cimento {\bf C18},(1995)517\par
 Ryazhskaya O.G., Nuovo Cimento {\bf C19},(1996)655

\bibitem{Mitsui}  Mitsui,K., Kitamura,T., Wada,T and Okei,K.,
 J.Phys. G: Nucl.Part.Phys.{\bf 29},(2003)2281

\bibitem{Sakai} Sakai,A., Ph.D thesis, University of Tokyo(1997)

\bibitem{Kasuga3} Kasuga,S., Ph.D thesis, University of Tokyo(1998)

\bibitem{GEANT} GEANT Detector Description and Simulation Tool,
    CERN Program Library W5013, CERN, Geneva (1994)

\bibitem{Fukunaga} K. Fukunaga, Introduction to Statistical Pattern
    Recognition, Academic, New York, (1972) 57

\bibitem{Kibayashi} Kibayashi,A., Ph.D thesis, University of Hawaii (2002)

\bibitem{Anokhina} Anokhina,A.M. and Galkin,V.I.,Physics of Atomic Nuclei {\bf 69} No.1 (2006) 16 

%
%
%
%
%
%
%
%
%
%


\end{thebibliography}
%
%
%

\end{document}